\documentstyle[preprint,aps,epsf]{revtex}

\def\singlespace {\smallskipamount=3.75pt plus1pt minus1pt
                  \medskipamount=7.5pt plus2pt minus2pt
                  \bigskipamount=15pt plus4pt minus4pt
                  \normalbaselineskip=12pt plus0pt minus0pt
                  \normallineskip=1pt
                  \normallineskiplimit=0pt
                  \jot=3.75pt
                  {\def\smallskip {\vskip\smallskipamount}}
                  {\def\medskip   {\vskip\medskipamount}}
                  {\def\bigskip   {\vskip\bigskipamount}}
                  {\setbox\strutbox=\hbox{\vrule
                    height10.5pt depth4.5pt width 0pt}}
                  \parskip 7.5pt
                  \normalbaselines}
\def\middlespace {\smallskipamount=5.625pt plus1.5pt minus1.5pt
                  \medskipamount=11.25pt plus3pt minus3pt
                  \bigskipamount=22.5pt plus6pt minus6pt
                  \normalbaselineskip=22.5pt plus0pt minus0pt
                  \normallineskip=1pt
                  \normallineskiplimit=0pt
                  \jot=5.625pt
                  {\def\smallskip {\vskip\smallskipamount}}
                  {\def\medskip   {\vskip\medskipamount}}
                  {\def\bigskip   {\vskip\bigskipamount}}
                  {\setbox\strutbox=\hbox{\vrule
                    height15.75pt depth6.75pt width 0pt}}
                  \parskip 11.25pt
                  \normalbaselines}
\def\doublespace {\smallskipamount=7.5pt plus2pt minus2pt
                  \medskipamount=15pt plus4pt minus4pt
                  \bigskipamount=30pt plus8pt minus8pt
                  \normalbaselineskip=30pt plus0pt minus0pt
                  \normallineskip=2pt
                  \normallineskiplimit=0pt
                  \jot=7.5pt
                  {\def\smallskip {\vskip\smallskipamount}}
                  {\def\medskip   {\vskip\medskipamount}}
                  {\def\bigskip   {\vskip\bigskipamount}}
                  {\setbox\strutbox=\hbox{\vrule
                    height21.0pt depth9.0pt width 0pt}}
                  \parskip 15.0pt
                  \normalbaselines}
\newcommand{\di}{\mathcal}

\tightenlines
\begin{document}
\preprint{
\hfill$\vcenter{\hbox{\bf IUHET-444} \hbox{September
             2001}}$  }

\title{\vspace*{.75in}
Leptogenesis and Low-energy Observables}

\author{M. S. Berger
\footnote{Electronic address:
berger@gluon.physics.indiana.edu}
and Kim Siyeon
\footnote{Electronic address:
siyeon@gluon.physics.indiana.edu}}

\address{
Physics Department, Indiana University, Bloomington, IN 47405, USA}

\maketitle

\thispagestyle{empty}

\begin{abstract}
We relate leptogenesis in a class of theories to low-energy 
experimental observables: quark and lepton masses and mixings.
With reasonable assumptions motivated by grand unification, one 
can show that the CP-asymmetry parameter takes a universal form. 
Furthermore the dilution mass is related to the light neutrino 
masses. Overall, these models offer an explanation for 
a lepton asymmetry in the early universe.
\end{abstract}

\newcommand{\be}{\begin{equation}}
\newcommand{\ee}{\end{equation}}
\newcommand{\bea}{\begin{eqnarray}}
\newcommand{\eea}{\end{eqnarray}}

\pacs{12.15.Ff, 14.60.Pq, 13.35.Hb}

\newpage

\section{Introduction}
Recent compelling evidence for neutrino oscillations has accelerated work on
formulating theoretical models for fermion masses and mixings. 
The current data indicates that there are most likely
two large mixing angles and one small one in the lepton sector. 
The first large mixing angle
arises in the atmospheric neutrino data, while it 
is becoming increasingly likely that the solar neutrino data is described by
an Mikheyev-Smirnov-Wolfenstein-(MSW) 
type oscillation with a large mixing angle 
(LMA)\cite{Fukuda:2001nj,Fukuda:2001nk,Toshito:2001dk,Ahmad:2001an}.
On the other hand the CHOOZ experiment\cite{Apollonio:1999ae}
gives an upper bound on the third mixing angle. 
A best fit\cite{Gonzalez-Garcia:2001ae}
for the atmospheric neutrino data and the LMA solution
for the solar neutrino data is
\begin{eqnarray}
&        \Delta m^2_{32}  =  3.2 \times 10^{-3}\ {\rm eV^2}\;,
\label{eq:mixing1}\\[2pt]
&        \sin^2 2\theta_{23}  =  1.000\;,
\label{eq:mixing2}\\[4pt]
&        \Delta m^2_{21} = 3.2 \times 10^{-5}\ {\rm eV^2}\;,
\label{eq:mixing3}\\[2pt]
&         \sin^2 2\theta_{12} = 0.75,\ \tan^2 \theta_{12} = 0.33\;,
\label{eq:mixing4}
\end{eqnarray}

The observations of neutrino mixing and the measured values for the differences
in mass-squareds make 
very plausible the existence of heavy Majorana neutrinos, $\nu_{M_i}$ .
These neutrinos can naturally be very heavy since they are Standard Model
gauge singlets and their masses are not connected to the breaking of the 
electroweak symmetry. These heavy Majorana neutrinos existed in the early 
universe and can have CP violating decay modes. Therefore 
the heavy neutrinos are natural candidates for producing
a lepton asymmetry via out-of-equilibrium 
decays. This asymmetry produced  in the early universe is recycled into 
a baryon asymmetry by sphaleron transitions which violated both baryon
number and lepton number. The resulting baryon asymmetry is the same order 
of magnitude as the original lepton asymmetry\cite{Buchmuller:2001sr}.

In the mass basis where the right-handed Majorana mass matrix $M_R$ is 
diagonal the asymmetry in heavy neutrino decays   
\bea
\epsilon_i&=&{{\Gamma (\nu_{M_i}\to \ell H_2)-\Gamma (\nu_{M_i}
\to \ell ^c H_2^c)}\over
{\Gamma (\nu_{M_i}\to \ell H_2)+\Gamma (\nu_{M_i}\to \ell ^c H_2^c)}}\;,
\eea
is given by\cite{Covi:1996wh,Buchmuller:1996pa,Flanz:1996fb,Buchmuller:1998yu,Rangarajan:1998hj,Pilaftsis:1999pd}
\bea
\epsilon_i&=&{3\over {16\pi v_2^2}}{1\over {({\di N}^\dagger {\di N})_{ii}}}
\sum_{n\ne i}{\rm Im} \left [({\di N}^\dagger {\di N})_{ni}^2\right ]
{{M_i}\over {M_n}}\;,\label{epsilon}
\eea
where $\di N$ is the neutrino Dirac mass matrix in a weak basis.
The masses $M_i$ are the
three eigenvalues of the heavy Majorana mass matrix and $v_2$ is
the vacuum expectation value (VEV)
of the Higgs boson giving Dirac masses to the neutrinos and up-type 
quarks. $M_1$ is the
mass of the lightest of the three heavy Majorana neutrinos, and 
Eq.~(\ref{epsilon}) is an approximate formula 
valid for $M_n >> M_i$. When this is the case,
the lepton asymmetry is generated
by the decays of the lightest Majorana neutrino, $\nu_{M_1}$.

The size of the lepton asymmetry generated by $\nu_{M_1}$ decays is also 
strongly dependent on the size of a mass parameter sometimes called the 
dilution mass defined as
\bea
\tilde{m}_1&=&{{({\di N}^\dagger {\di N})_{11}}\over {M_1}}\;. \label{m1t}
\eea
This parameter controls (a) the decay width of the lightest 
right-handed Majorana neutrino $\nu_{M_1}$
since
\bea
&&\Gamma _{\nu_{M_1}}
={1\over {8\pi}}({\di N}^\dagger {\di N})_{11}{{M_1}\over {v_2^2}}\;,
\eea
as well as (b) the amount of dilution caused by lepton number 
violating scattering: the resulting lepton asymmetry 
 depends critically on the 
parameter $\tilde{m}_1$ because it governs the size of the most important 
Yukawa coupling in the $\Delta L=2$ scattering processes, as has been shown 
in detail in numerical 
calculations\cite{Buchmuller:2001sr,Buchmuller:1996pa,Buchmuller:1998yu,Berger:1999bg,Berger:2000fv}.
These two constraints bound the possible values of
$\tilde{m}_1$ such that a sufficient asymmetry is produced to agree with 
observation. The generated lepton asymmetry $Y_L$ is defined in terms
of the number densities of the leptons and antileptons as well as the entropy 
density as 
\bea
&&Y_L={{n_L-n_{\overline L}}\over s}=\kappa {{\epsilon _1}\over g^*}\;,
\label{YL}
\eea
where $g^*$ is the number of light (effective) 
degrees of freedom in the theory, and  
$\kappa $ is a dilution 
factor that can be reliably calculated by solving the full Boltzmann 
equations. 

It has been shown\cite{Buchmuller:1996pa,Buchmuller:1998yu} that a 
CP-violation parameter $\epsilon_1\sim 10^{-6}$ and a dilution mass 
$\tilde{m}_1$ in the range of the light neutrino masses can produce the 
sufficient amount of leptogenesis to account for the observed baryon 
asymmetry. From the definition of the dilution mass
in Eq.~(\ref{m1t}) it is clear that the dilution mass will indeed be
related to the light neutrino masses in most models.
It is a nontrivial occurrence that the amount of 
baryon asymmetry of the universe is obtained from a recycling of the 
leptogenesis that naturally occurs via Majorana neutrino decays. 

Suppose one starts in a basis where $M_R$ is diagonal with 
eigenvalues $M_j$, and suppose the matrix $M_R$ is
connected to the light neutrino mass matrix $m_\nu$ by a seesaw mechanism,
\be
M_R = {\di N}^T m_\nu^{-1} {\di N}\;.
\label{seesaw}
\ee
One can then define mixing matrices $U_{L,R}^{(N)}$ and $V_L$ that diagonalize
$N$ and $m_\nu$ respectively,
\bea
&& {\di N} = U_L^{(N)} N_{diag} U_R^{(N) \dagger}\;,  \\
&& {\di N}^T = U_R^{(N) *} N_{diag} U_L^{(N) T}\;,  \\ 
&& m_\nu^{-1} = V_L^* m_{diag}^{-1} V_L^\dagger\;. 
\eea
With these transformation matrices defined, 
$M_j$ can be written in terms of mass eigenvalues and mixings of 
$(m_\nu, {\di N})$,
\bea
&&M_j = \sum_k \sum_\ell m_k^{-1} \left( V_L^\dagger U_L^{(N)} 
           \right)_{k\ell}^2 N_\ell^2 U^{(N) \dagger 2}_{R\ell j} \;
\label{mjay}
\eea
where $N_l$ are the diagonal elements of $N_{diag}$.
The unitary transformation $U_R^{(N)}$ diagonalizes ${\di N}^\dagger {\di N}$ 
as ${\di N}^\dagger {\di N} = U_R^{(N)} N_{diag}^2 U_R^{(N) \dagger}$, then
\bea
\left ({\di N}^\dagger {\di N}\right )_{1j} 
&=& \sum_k N_k^2 U_{R1k}^{(N)} U_{Rjk}^{(N)*} 
\eea

By inverting $M_j$ in Eq.~(\ref{mjay}), the mass eigenvalues of $m_\nu$ 
can be expressed in terms of $V_L$, mixing angles and eigenvalues of 
Dirac mass matrix, and Majorana neutrino masses, 
\bea
&&m_k = \sum_j \sum_\ell \left( V_L^\dagger U_L^{(N)} 
           \right)_{k\ell}^2 N_\ell^2 U^{(N) \dagger 2}_{R\ell j} M_j^{-1} \;
\label{smallmkay}
\eea

\section{Assumptions}
It is well-known that one must make theoretical assumptions about the structure
of the neutrino masses and mixings to make progress in ascertaining whether
leptogenesis is viable. For example the source of CP-violation responsible for
producing the CP-violating decays of heavy Majorana neutrinos (and hence giving
rise to leptogenesis) does not have to be related to the CP-violation that 
might be measurable at low-energy experiments in the 
future\cite{Plumacher:1998ex,Branco:2001pq}.
An extensive study of the weak-basis CP invariants in 
models with three iso-singlet neutrinos is given in 
Ref.~\cite{Pilaftsis:1997jf}.
If one makes the assumption of single right-handed neutrino dominance, then
the low energy neutrino observables and the leptogenesis predictions 
decouple entirely\cite{Hirsch:2001dg}. On the other hand, in certain
classes of grand unified theories previously unconstrained parameters become
related to observables. For example, in models with a left-right symmetry, 
the right-handed mixing angles can be related the left-handed ones that 
enter into low-energy experiments\cite{Falcone:2000mf}.
In this section we list our theoretical assumptions about the underlying
grand unified theory. Many authors have discussed leptogenesis in the context
of grand unified 
theories\cite{Mohapatra:1992pk,Falcone:2000ma,Falcone:2000eg,Falcone:2001im,Falcone:2001ib,Buccella:2001tq,Plumacher:1997kc,Plumacher:1998ru,Carlier:1999ac,Ma:1999sq,Goldberg:2000hp,Kang:2000fr,Nezri:2000pb,Joshipura:1999is,Joshipura:2001ui,Buchmuller:2001dc}; our emphasis 
here is on making the most
general assumptions that allow us to relate low-energy observables like
masses and mixing angles to the required lepton asymmetry that can ultimately
account for the baryon asymmetry of the universe.

{\bf [A1]} We assume that the Dirac mass matrices 
${\di N}$ and ${\di U}$ are symmetric, and 
${\di N}\sim {\di U}$\footnote{We use the notation $\sim$ 
to denote that entries are the 
same size to leading order
in all small quantities such as small mass ratios or small mixing angles.}. 
This similarity between the neutrino Dirac mass matrix and the up-quark 
mass matrix is motivated by grand unified theories.

{\bf [A2]} The mixing angles contained in the transformation matrices that 
diagonalize the neutrino Dirac mass matrix
${\di N}$ are related to the eigenvalues \footnote{We use the 
shorthand notation
$c_{ij}\equiv \cos \theta_{ij}$ and $s_{ij}\equiv \sin \theta_{ij}$.}
$s_{ij} \sim \sqrt{N_i \over N_j}$.
In general these mixing angles cannot be larger than $\sqrt{N_i \over N_j}$, 
but can in principle be smaller. The $s_{ij}$ being suppressed compared 
to $\sqrt{N_i \over N_j}$ might occur, for example, if some 
elements of ${\di N}$ are suppressed or zero. So the result of our second
assumption is that there is no such suppression or cancellation in the 
Dirac neutrino matrix.

The crucial features that follow from our two assumptions listed above
are (a) the 
neutrino Dirac mass matrix has eigenvalues that mimic the large hierarchy
that exists in the up-quark sector, and (b)  the mixing angles $s_{ij}$ are
fixed to be of some definite size related to the up-type quark masses, e.g. 
$s_{13}\sim \sqrt{N_1/N_3}\sim \sqrt{m_u/m_t}$. These two results will be 
important in arriving at the relatively simple results that follow.  

{\bf [A3]} Our approach does not allow us to determine the CP-violating phase
that enters into the parameter $\epsilon_1$ in Eq.~(\ref{epsilon}). We simply 
assume that phases are of order one, and there is no suppression arising from 
unnaturally small parameters.

A standard parametrization
of the unitary transformation involving
three angles and a phase is 
\bea 
U &=& \left(
\begin{array}{ccc}
1 & 0       & 0        \\
0 & c_{23}  & s_{23}   \\
0 & -s_{23} & c_{23}   \\
\end{array} \right) \left(
\begin{array}{ccc}
c_{13}              & 0 & s_{13} e^{-i\delta}  \\
0                   & 1 & 0                    \\
-s_{13} e^{i\delta} & 0 & c_{13}               \\
\end{array} \right) \left(
\begin{array}{ccc}
c_{12}  & s_{12} & 0  \\
-s_{12} & c_{12} & 0  \\
0       & 0      & 1  \\
\end{array} \right) \cr
&& \cr 
&=& \left(
\begin{array}{ccc}
c_{13}c_{12} &  s_{12}c_{13} & s_{13}e^{-i\delta} \\
-s_{12}c_{23}-s_{23}s_{13}c_{12}e^{i\delta} & 
c_{23}c_{12}-s_{23}s_{13}s_{12}e^{i\delta} & s_{23}c_{13} \\
s_{23}s_{12}-s_{13}c_{23}c_{12}e^{i\delta} & 
-s_{23}c_{12}-s_{13}s_{12}c_{23} & c_{23}c_{13} \\
\end{array} \right)\;.
\eea
The right-handed and left-handed mixing matrices with small angles 
($c_{ij}\approx 1$) are
\be
U_R^{(N)} \approx U_L^{(N)} \approx \left(
\begin{array}{ccc}
1 &  s_{12} & s_{13} \\
-s_{12}-s_{23}s_{13} & 1 & s_{23} \\
s_{23}s_{12}-s_{13} & -s_{23}-s_{13}s_{12} & 1 \\
\end{array} \right)\;,
\ee
where we will assume that phase $e^{i\delta }$ is not suppressed: $\delta$ is
not close to $0$ or $\pi$.
For our purposes, we consider only the leading contributions to each element
so that
\be
|U_R^{(N)}| \sim |U_L^{(N)}| \sim \left(
\begin{array}{ccc}
1 &  s_{12} & s_{13} \\
s_{12} & 1 & s_{23} \\
s_{13} & s_{23} & 1 \\
\end{array} \right)
\label{Nmixing}
\ee

\section{Neutrino transformation}
In general, we can write the transformation as 
\bea 
&&V_L \sim \left(
\begin{array}{ccc}
1 & \Theta_{12}      & \Theta_{13}        \\
-\Theta_{12}-\Theta_{23}\Theta_{13} & 1  & \Theta_{23}   \\
\Theta_{23}\Theta_{12}-\Theta_{13} & -\Theta_{23}-\Theta_{13}\Theta_{12} & 1 \\
\end{array} \right) \;.
\label{VL}
\eea
We henceforth interpret the quantities $\Theta _{ij}$ as
\begin{eqnarray}
&&\cos\Theta \sim 1\;, \quad \sin\Theta \sim 1\;, \quad {\rm for \:\:
large\:\: angles}
\nonumber \\
&&\cos\Theta \sim 1\;, \quad \sin\Theta \sim \Theta\;, \quad 
{\rm for \:\: small \:\: angles}.
\label{Theta}
\end{eqnarray}
In other words, the matrix can be expressed in the same way in 
terms of $\Theta _{ij}$ if we are only interested in the order-of-magnitude 
size of the elements (including the ones on the diagonal which would only be of
order one in general). 
The Maki-Nakagawa-Sakata (MNS) neutrino mixing matrix\cite{Maki:1962mu} is 
\begin{eqnarray}
&&U_{MNS}=U_L^{(E)\dagger}V_L\;,
\label{mns}
\end{eqnarray}
where $U_L^{(E)}$ is the matrix that diagonalizes the charged lepton mass
matrix.
The constraints from reactor neutrino mixing data\cite{Apollonio:1999ae}
imply that $\Theta _{13}$ 
must be small provided there is no cancellation among $V_L$ and 
$U_L^{(E)}$. Retaining only information about the size of the 
individual elements, we may write Eq.~(\ref{VL}) as 
follows
\bea 
&&V_L \sim \left(
\begin{array}{ccc}
1 & \Theta_{12}      & \Theta_{13}        \\
{\rm max}(\Theta_{12},\Theta_{23}\Theta_{13}) & 1  & \Theta_{23}   \\
{\rm max}(\Theta_{23}\Theta_{12},\Theta_{13}) & \Theta_{23} & 1 \\
\end{array} \right) \;,
\label{theta}
\eea
with the entries interpreted according to Eq.~(\ref{Theta}).

\section{Heavy Majorana neutrino masses}

Define the matrix 
\begin{eqnarray}
&&W_{kj}\equiv \sum _{\ell} \left (V_L^\dagger U_L^{(N)}
\right )^2_{k\ell}n^2_\ell U_{R\ell j}^{(N)\dagger 2}\;,
\end{eqnarray}
where $n_i\equiv N_i/N_3$ are the ratios of the Dirac neutrino masses.
The heavy Majorana neutrino masses are 
\begin{eqnarray}
&&M_j=N_3^2\sum _k m_k^{-1}W_{kj}\;,
\end{eqnarray}
and the light neutrino masses are given by Eq.~(\ref{smallmkay}) as  
\begin{eqnarray}
&&m_j=N_3^2\sum _k W_{jk}M_k^{-1} \;.
\label{lightmjay2}
\end{eqnarray}
The factor $n^2_\ell U_{R\ell j}^{(N)\dagger 2}$ in $W_{kj}$ has the form 
\begin{eqnarray}
&&n^2_\ell U_{R\ell j}^{(N)\dagger 2}\sim 
\left(
\begin{array}{ccc}
n_1^2 & n_1^2s_{12}^2 & n_1^2s_{13}^2        \\
n_2^2s_{12}^2 & n_2^2 & n_2^2s_{23}^2  \\
s_{13}^2 & s_{23}^2 & 1 \\
\end{array} \right) \;.
\end{eqnarray}
Now we make use of our assumptions [A1] and [A2] 
that allow us to compare
the relative sizes of the $n_i$ and the mixing angles $s_{ij}$. Specifically
we have that $s_{ij}\sim \sqrt{n_i/n_j}$ as well as $n_i<<n_j$ for $i<j$ 
so that
\begin{eqnarray}
&&n_1^2U_{R1j}^{(N)\dagger 2}<<n_2^2U_{R2j}^{(N)\dagger 2}
<<U_{R3j}^{(N)\dagger 2}\;.
\label{hierarchy}
\end{eqnarray}
We henceforth refer to this condition as  
``third-generation dominance.''
In fact if, as we have assumed, the hierarchy in the Dirac masses for neutrinos
is as strong as it is for the up quark Dirac masses, as one might expect in a 
grand unified theory, then the smallness of $n_1$ and $n_2$ suppresses 
all other
contributions to $W_{kj}$ relative to the dominant contribution coming from 
$\left (V_L^\dagger U_L^{(N)}\right )^2_{k3}$ and 
$U_{R3j}^{(N)\dagger 2}$. So we arrive at the following factored form
for the matrix
\begin{eqnarray}
&&W_{kj}\sim \left (\begin{array}{c}
(V_L^\dagger U_L^{(N)})^2_{13} \\
(V_L^\dagger U_L^{(N)})^2_{23} \\
(V_L^\dagger U_L^{(N)})^2_{33} 
\end{array} \right)
\left ( \begin{array}{ccc}
s_{13}^2 & s_{23}^2 & 1
\end{array} \right)\;.
\end{eqnarray}
Finally we can write the Majorana masses in the following way
\begin{eqnarray}
&&\left (M_1,M_2,M_3\right )\sim N_3^2\tilde{W}_3
\left (s_{13}^2,s_{23}^2,1\right )\;,
\label{massrat}
\end{eqnarray}
where 
\begin{eqnarray}
&&\tilde{W}_3=\sum_k m_k^{-1}(V_L^\dagger U_L^{(N)})^2_{k3}\;.
\label{W3}
\end{eqnarray}
The result in Eq.~({\ref{massrat}) 
indicates that, based on our assumptions, the mass ratios of the 
Majorana masses are related to the mixing angles $s_{i3}$ and are 
independent of the light neutrino mixings which appear only in the overall
factor $\tilde{W}_3$. This result follows from the third-generation dominance 
Eq.~(\ref{hierarchy}) which is related to the large hierarchy in the 
Dirac neutrino masses that is inherited from the large hierarchy in the 
experimentally measured up-quark masses. 
On the other hand, the light neutrino masses under the third-generation
condition are given by Eq.~(\ref{lightmjay2}) as
\begin{eqnarray}
&&\left (m_1,m_2,m_3\right )\sim {N_3^2 \over M_3} 
\left ((V_L^\dagger U_L^{(N)})^2_{13},
(V_L^\dagger U_L^{(N)})^2_{23},
(V_L^\dagger U_L^{(N)})^2_{33}\right )\;.
\label{massratio}
\end{eqnarray}
So the mass ratios of the light neutrinos can be expressed in terms of the 
left-handed mixing angles.
 
\section{Leptogenesis}
In this section we utilize the simple form for the mass ratios of the heavy 
Majorana neutrino masses found in the last section to derive a simple formula
for the CP-asymmetry parameter $\epsilon _1$ in Eq.~(\ref{epsilon}).
The couplings give
\begin{eqnarray}
&&\left ({\di N}^\dagger {\di N}\right )_{1j}
=N_3^2\sum_k n_k^2U_{R1k}^{(N)} U_{Rjk}^{(N)*}\;,
\end{eqnarray}
where the dominant contribution is given in this case by $k=3$,
\begin{eqnarray}
\left ({\di N}^\dagger {\di N}\right )_{1j}
\sim N_3^2 U_{R13}^{(N)} U_{Rj3}^{(N)*}\;.
\end{eqnarray}
As with the Majorana masses, third-generation dominance implies that 
simple expressions exist for 
\begin{eqnarray}
&&\left [({\di N}^\dagger {\di N})_{11},({\di N}^\dagger {\di N})_{12},
({\di N}^\dagger {\di N})_{13}\right ]
\sim N_3^2s_{13}\left [s_{13},s_{23},1\right ]\;.
\label{couprat}
\end{eqnarray}
The resulting CP-asymmetry parameter in Eq.~(\ref{epsilon}) can now be 
expressed to leading order as
\begin{eqnarray}
&&\epsilon _1\sim {3\over {16\pi}}{{N_3^2}\over {v_2^2}}{\rm Im}\left [
{{{({\di N}^\dagger {\di N}})_{12}^2}\over 
{{({\di N}^\dagger {\di N}})_{11}}}{M_1\over M_2}+
{{{({\di N}^\dagger {\di N}})_{13}^2}\over 
{{({\di N}^\dagger {\di N}})_{11}}}{M_1\over M_3}\right ]\;.
\label{epsilon_mid}
\end{eqnarray}
and one arrives at the simple result
\begin{eqnarray}
&&\epsilon _1\sim 10^{-1}s_{13}^{2}\sim 10^{-1}{{m_u}\over {m_t}}\;,
\label{epsilon_a}
\end{eqnarray}
where we have used Eqs.~(\ref{massrat}) and (\ref{couprat}).
We have also used $N_3 \sim v_2$ since the largest Yukawa coupling
in the neutrino Dirac mass matrix is similar (${\di N}\sim {\di U}$)
to the top quark Yukawa coupling
which is close to one.
One can understand that the contribution involving the mixing angle
$s_{13}^{2}$ is the leading contribution in the following way: The dominant
contribution to leptogenesis comes from the decay of the lightest 
Majorana neutrino ($i=1$) and the dominant Yukawa couplings occur in the 
third generation ($j=3$). One obtains an acceptable amount of baryon asymmetry
if $\epsilon _1\sim 10^{-6}$; this indeed results if 
$s_{13}\sim \sqrt{m_u/m_t}$.

The dilution mass defined in Eq.~(\ref{m1t}) can be expressed as
\begin{eqnarray}
&&\tilde{m}_1\sim {{N_3^2s_{13}^2}\over {N_3^2\tilde{W}_3s_{13}^2}}
=\tilde{W}_3^{-1}\;,
\label{dilute}
\end{eqnarray}
using the third-generation-dominance that results from assumptions
[A1] and [A2], 
Given the expression for $\tilde{W}_3$ in Eq.~(\ref{W3}) it is clear that 
the dilution mass is related in all cases to the light neutrino masses.
This is precisely the range of dilution mass that gives a large asymmetry as
has been pointed out many times before as an attractive and natural feature
of the leptogenesis scenario.
 
We now proceed to examine some special cases for the size of the 
dilution mass. Assumptions [A1] and [A2] allow us to identify 
the sizes of the mixing angles in the the mixing matrix $U_L^{(N)}$.
For example $s_{23}\sim \sqrt{m_c/m_t}$. 
So Eq.~(\ref{Nmixing}) can be written as  
\begin{eqnarray}
&&U_L^{(N)}\sim\left( \begin{array}{ccc}
1 & s_{12} & s_{13} \\
s_{12}& 1 & s_{23} \\
s_{13} & s_{23} & 1 \\
\end{array} \right)
\end{eqnarray}
Recall that the left-handed mixing angles are similar to the right-handed 
mixing angles according to our assumptions.
Using Eq.~(\ref{theta}) we have that
\begin{eqnarray}
&&(V_L^\dagger U_L^{(N)})^2_{k3}\sim
\left( \begin{array}{c}
{\rm max}(s_{13}^2,\Theta_{12}^2s_{23}^2,\Theta_{23}^2\Theta _{12}^2,
\Theta_{23}^2\Theta_{13}^2s_{23}^2,\Theta_{13}^2) \\
{\rm max}(s_{23}^2,\Theta_{23}^2) \\
1 \\
\end{array} \right)\;.
\label{vu}
\end{eqnarray}
These elements together with Eqs.~(\ref{W3}) and (\ref{dilute}) allow 
one to determine the dilution mass. The quantities $s_{ij}$ are all small 
compared with one since they have been related to the (left-handed) 
Cabibbo-Kobayashi-Maskawa (CKM) matrix elements, 
but the $\Theta _{ij}$ might or might not be small. From the CHOOZ 
data\cite{Apollonio:1999ae} we 
know that the mixing angle $\Theta _{13}$ must be small as long as there
is no unnatural cancellation between this angle and the 
one involved in converting the weak basis to the mass basis for the charged 
leptons, c.f. Eq.~(\ref{mns}). 
One can relate the dilution mass in Eq.~(\ref{dilute}) to the light
neutrino masses using Eqs.~(\ref{W3}), (\ref{massratio}), and (\ref{vu}). 
The mass ratios between light neutrinos are
\begin{eqnarray}
&& {m_i \over m_j} \sim 
{(V_L^\dagger U_L^{(N)})^2_{i3} \over
(V_L^\dagger U_L^{(N)})^2_{j3}}\;.
\label{mixratio}
\end{eqnarray}
One can investigate a number of cases. Without any fine-tuning one expects
the angles $\Theta_{ij}$ to be of the same order as the angles 
$s_{ij}$. In that case,
1) one obtains 
\begin{eqnarray}
m_2\sim {{m_c}\over {m_t}}m_3\;, \qquad m_1\sim {{m_u}\over {m_t}}m_3\;, 
\end{eqnarray}
from Eqs.~(\ref{vu}) and (\ref{mixratio}). The dilution mass is
$\tilde{m}_1\sim m_3$ from Eqs.~(\ref{W3}) and (\ref{dilute}).
This does not give good 
agreement with the experimental data since $m_2\simeq \sqrt{\Delta m_{21}^2}$ 
is too small to reconcile
it with the solar LMA data and atmospheric data neutrino data 
$m_3\simeq \sqrt{\Delta m_{32}^2}$ .
The neutrino masses inherit the large hierarchy from the up quark sector.
The conclusion is that one needs some amount of fine-tuning to get masses 
in acceptable agreement with the solar LMA data.

2) If one accepts some fine-tuning so that the mixing angle $\Theta _{23}$ 
is large and order one rather than similar to $s_{23}$ and $\Theta _{12}$ 
remains small, then
Eq.~(\ref{vu}) reduces to 
\begin{eqnarray}
&&(V_L^\dagger U_L^{(N)})^2_{j3}\sim
\left( \begin{array}{c}
\Theta_{12}^2 \\
1 \\
1 \\
\end{array} \right)\;.
\end{eqnarray}
Even in this case the determinant of the seesaw mass formula, 
Eq.~(\ref{seesaw}), must satisfy
\begin{eqnarray}
&&m_1m_2m_3=m_um_cm_t\left ({{m_t}\over {M_3}}\right )^3\;.
\label{determinant}
\end{eqnarray}
Then since $\Theta _{23}$ is large one expects the mass eigenvalues to 
satisfy
\begin{eqnarray}
m_2 \sim m_3\sim \sqrt{m_cm_t}\left ({{m_t}\over {M_3}}\right )\;,
\end{eqnarray}
so that 
\begin{eqnarray}
m_1 \sim m_u\left ({{m_t}\over {M_3}}\right )\;.
\end{eqnarray}
The masses can be consistent with the LMA solar and atmospheric neutrino data.
Then the dilution mass is given by $\tilde{m}_1\sim m_2\sim m_3$ from
Eqs.~(\ref{W3}) and (\ref{dilute}) and is in an acceptable range. 

3) With additional fine-tuning both the mixing angles 
$\Theta _{12}$ and $\Theta _{23}$ can be made large. 
Eq.~(\ref{vu}) reduces to 
\begin{eqnarray}
&&(V_L^\dagger U_L^{(N)})^2_{j3}\sim
\left( \begin{array}{c}
1 \\ 1 \\ 1 \\
\end{array} \right)\;.
\end{eqnarray}
The light neutrino masses are all the same order so that from 
Eq.~(\ref{determinant}) one gets
\begin{equation}
m_1 \sim m_2 \sim m_3 \sim (m_um_cm_t)^{1/3}
\left ({{m_t}\over {M_3}}\right )\;,
\end{equation}
where $m_i$ cannot be larger than $m_j$ if $i<j$.
Then the dilution mass is $\tilde{m}_1\sim m_1$.
This solution does not offer any explanation for a hierarchy in neutrino 
masses.

In all three cases the 
dilution mass $\tilde{m}_1$ lies roughly in the range spanned by
light neutrino masses
\begin{eqnarray}
m_1\alt \tilde{m}_1 \alt m_3\;.
\end{eqnarray}
It should be understood here that the $\alt$ means that $\tilde{m}_1$ could 
be outside the upper and lower ends of the range by an order one
parameter.

More generally, and outside the assumptions of this paper, one can
consider the possibility that the charged lepton mass matrix contributes  
to large mixing for both the solar neutrino and atmospheric netrino 
oscillations or for either one, 
though the charged lepton transformation 
matrix $U_L^{(E)}$ via Eq.~(\ref{mns}).

\section{Summary}
We have shown that based upon a limited number of reasonable assumptions
about the neutrino sector motivated by grand unification, 
one obtains the universal expression in 
Eq.~(\ref{epsilon_a}) for the dominant contribution to the  
CP-violation parameter $\epsilon _1$ that determines the amount of 
leptogenesis in the early universe. Furthermore the dilution mass $\tilde{m}_1$
is expressed in terms of mixing angles in the light neutrino masses and
it naturally falls in the range needed to explain the baryon asymmetry of the
universe. While these assumptions are not 
required to obtain the necessary lepton
asymmetry to explain the observed baryon asymmetry of the universe, they 
provide enough constraints to allow one to relate the CP-violation in 
the heavy Majorana neutrino decays and the important Yukawa couplings 
of these heavy neutrinos to low-energy observables: fermion masses and mixing
angles.


\vspace{0.5cm}

\section*{Acknowledgments}

This work was supported in part by the U.S.
Department of Energy under Grant No. No.~DE-FG02-91ER40661.



\end{document}